\documentclass[conference]{IEEEtran}
\IEEEoverridecommandlockouts

\usepackage{cite}
\usepackage{amsmath,amssymb,amsfonts}
\usepackage{graphicx}
\usepackage{textcomp}
\usepackage{xcolor}
\usepackage{booktabs}
\usepackage{url}
\usepackage{stfloats}
\usepackage{tikz}
\usepackage{flushend}

\begin{document}
\bstctlcite{BSTcontrol}

\title{Architecture Without Architects: How AI Coding Agents Shape Software Architecture}

\author{
  \IEEEauthorblockN{Phongsakon Mark Konrad}
  \IEEEauthorblockA{\textit{Centre for Industrial Software}\\
    \textit{University of Southern Denmark}\\
    Alsion 2, S{\o}nderborg, 6400, Denmark\\
  phkon23@student.sdu.dk}
  \and
  \IEEEauthorblockN{Tim Lukas Adam}
  \IEEEauthorblockA{\textit{Centre for Industrial Software}\\
    \textit{University of Southern Denmark}\\
    Alsion 2, S{\o}nderborg, 6400, Denmark\\
  tiada23@student.sdu.dk}
  \and
  \IEEEauthorblockN{Riccardo Terrenzi}
  \IEEEauthorblockA{\textit{Centre for Industrial Software}\\
    \textit{University of Southern Denmark}\\
    Alsion 2, S{\o}nderborg, 6400, Denmark\\
  rite@mmmi.sdu.dk}
  \and
  \IEEEauthorblockN{Serkan Ayvaz}
  \IEEEauthorblockA{\textit{Centre for Industrial Software}\\
    \textit{University of Southern Denmark}\\
    Alsion 2, S{\o}nderborg, 6400, Denmark\\
  seay@mmmi.sdu.dk}
}

\maketitle

\begin{abstract}
AI coding agents select frameworks, scaffold infrastructure, and wire integrations, often in seconds. These are architectural decisions, yet almost no one reviews them as such. We identify five mechanisms by which agents make implicit architectural choices and propose six prompt-architecture coupling patterns that map natural-language prompt features to the infrastructure they require. The patterns range from contingent couplings (structured output validation) that may weaken as models improve to fundamental ones (tool-call orchestration) that persist regardless of model capability. An illustrative demonstration confirms that prompt wording alone produces structurally different systems for the same task. We term the phenomenon \textit{vibe architecting}, architecture shaped by prompts rather than deliberate design, and outline review practices, decision records, and tooling to bring these hidden decisions under governance.

\end{abstract}


%% ============================================================
\section{Introduction}
\label{sec:intro}

A growing number of developers build software by describing what they want in plain language, a practice Karpathy calls \textit{vibe coding}~\cite{b1}. What started as line-level autocomplete in 2022 has grown into agents that produce entire systems from a single description.

But these tools do far more than write code. Claude Code~\cite{b2} scaffolds full projects and delegates subtasks to sub-agents. Cursor~\cite{b3} runs background agents across parallel worktrees. Devin~\cite{b4} supports interactive planning. Bolt.new~\cite{b5} produces full-stack applications in browser containers. Codex~\cite{b6} runs in sandboxed cloud environments. Windsurf~\cite{b7} adjusts its output dynamically. In each case, they pick frameworks, configure databases, and set up authentication. Each of those choices is an architectural decision.

Two phenomena are the focus of this paper. The first is \textit{agent-driven architecture}, in which agents choose frameworks, databases, and deployment targets without justifying their choices. The second is \textit{prompt-architecture coupling}, in which Large Language Model (LLM)-integrated code declares tool access, and something must orchestrate those calls. When it injects retrieved documents, a retrieval pipeline follows. In both cases, the prompt dictates the infrastructure. Together, these produce what is here termed \textit{vibe architecting}, architecture shaped by natural-language prompts rather than by deliberate, recorded design.

Architects need ways to catch and guide these decisions. This position paper combines a tool survey with an illustrative case study and makes three contributions:
\begin{enumerate}
    \item Five mechanisms are identified by which agents make implicit architectural decisions, illustrated by a demonstration where prompt wording is the sole varying factor.
    \item Six prompt-architecture coupling patterns are proposed that map prompt-level features to the infrastructure they require, drawing on LangChain~\cite{b8}, LlamaIndex~\cite{b9}, and provider documentation.
    \item Several open questions are raised for the architecture community, including converging stacks, new forms of technical debt, and a widening gap between how fast agents build and how fast teams can review.
\end{enumerate}

%% ============================================================
\section{Background and Related Work}
\label{sec:related}

White~\textit{et~al.}~\cite{b10} catalog prompt patterns, drawing an analogy to classical design patterns. Chen~\textit{et~al.}~\cite{b11} go further and treat prompts as first-class software artifacts (``promptware'') with lifecycle concerns including versioning, testing, and deployment. Neither work investigated what happens when those prompt choices start dictating the architecture of the surrounding system.

LangChain~\cite{b8} and LlamaIndex~\cite{b9} organize their APIs around architectural abstractions (chains, agents, retrievers), but that organization lives inside library code rather than in reusable design knowledge a team could reason about on its own. Wang~\textit{et~al.}~\cite{b12} survey LLM-based autonomous agents; Wu~\textit{et~al.}~\cite{b13} propose multi-agent conversation frameworks. Neither inspects how a prompt's structure implies particular components. In a systematic literature review of software architecture and LLMs, Schmid~\textit{et~al.}~\cite{b14} found growing work applying LLMs to architecture tasks but very little asking how AI tools shape the architecture of the systems they help build.

Sculley~\textit{et~al.}~\cite{b15} remains the classic reference on hidden debt in ML systems, namely configuration debt, data dependency debt, and feedback loops. Al~Mujahid and Imran~\cite{b16} studied GenAI-induced self-admitted technical debt, documenting cases where developers pull in AI-generated code while openly admitting that they are unsure whether it works. Kravchuk-Kirilyuk~\textit{et~al.}~\cite{b17} focused on modularity and showed that LLM-generated code often appears modular on the surface while violating deeper modular principles. The concern here is different. Sculley~\textit{et~al.}'s~\cite{b15} debt accumulates during training and deployment. The debt identified in this paper accumulates during code generation, before the system even runs. 

Sarkar and Drosos~\cite{b18} characterize vibe coding as ``programming through conversation''; Fawzy~\textit{et~al.}~\cite{b19} reviewed practitioner accounts of how developers use these workflows and Falc{\~a}o~\textit{et~al.}~\cite{b20} place AI-driven development on the longer trajectory toward self-coding systems. All three focus on developer interaction and adoption. None examines architectural consequences. The following section identifies the specific mechanisms through which those consequences arise. 

%% ============================================================
\section{Agentic Coding as Architectural Decision-Making}
\label{sec:agentic}

AI coding agents are making architectural decisions, and recent adoption data~\cite{b21} suggests this is a progressive phenomenon rather than a transient one. The scope of what agents decide has widened rapidly. Line-level autocomplete (2022) had almost no architectural footprint~\cite{b22}. By 2024, multi-file tools~\cite{b3,b7} were editing across files, indirectly drawing module boundaries. System-level agents arrived in 2025 (Claude Code~\cite{b2}, Devin~\cite{b4}, Codex~\cite{b6}, Kiro~\cite{b23}, Replit Agent~\cite{b24}, among others), and can scaffold entire projects from a single prompt. More recently, coordinated agents split work across sub-agents assigned to specific files, turning decomposition itself into an architectural concern~\cite{b25}. By 2026, the agent's decision surface covers the full breadth of software architecture as defined by Bass~\textit{et~al.}~\cite{b25}. Standardization through MCP~\cite{b26} and the \texttt{AGENTS.md} convention~\cite{b27} has reinforced this trend, forming a de facto integration architecture. Yet these decisions take seconds, arrive bundled, and \textit{leave no record}.

\subsection{Mechanisms and Evolution}

Five mechanisms were identified through which current agents make architectural decisions. They were derived by surveying six tools (Table~\ref{tab:mechanisms}) illustrates how they manifest across three prompt variants.

\textit{1)~Model selection.} Different LLMs produce structurally different code. SonarSource found that each model exhibits a distinct ``coding personality''~\cite{b28}. Switching a model selector is an architectural choice, even if nobody frames it that way.

\textit{2)~Task decomposition.} How agents split work shapes architecture. Claude Code delegates subtasks to sub-agents working on individual files; Cursor's background agents operate in parallel worktrees. Because decomposition determines module boundaries~\cite{b25}, the agent designs the system's modular structure.

\textit{3)~Default configuration.} Guardrail mechanisms exist (Claude Code hooks, Cursor's \texttt{.cursorrules}, \texttt{AGENTS.md}~\cite{b27}), but all require the developer to set them up. Without explicit rules, agents default to training-data priors.

\textit{4)~Scaffolding and autonomous generation.} Templates pre-select framework, database, authentication, and deployment; the architectural choice was made when the template was written. Autonomous agents go further: a single prompt can produce a complete project, folding every choice into one interaction with no visible rationale.

\textit{5)~Integration protocols.} Tools built on MCP~\cite{b26} produce systems whose integration points follow a standard protocol for tool discovery, invocation, and data exchange. Older tools hard-code service access, binding the system to the generation platform. Either way, the integration architecture is chosen by the tool, not by the team.

To demonstrate these mechanisms, three variants of a customer-service chatbot were built.\footnote{Source code: \url{https://github.com/phomarkon/vibe-architecting-case-study}} Each was developed independently with no shared context. The architectural choices that emerged were recorded (Table~\ref{tab:mechanisms}). All three variants were generated using Claude Code with default settings; the exact prompts and full generation details are documented in the companion repository. All variants ran on the same LLM at runtime (GPT-4o-mini), so prompt specification was the sole varying factor. The case study illustrates prompt-architecture coupling (Section~\ref{sec:coupling}); the five mechanisms themselves are supported independently by the tool survey above. The intent is illustration, not experimental proof.

\begin{table}[htbp]
  \caption{Architectural Divergence Across Three Prompt Variants}
  \label{tab:mechanisms}
  \centering
  \scriptsize
  \begin{tabular}{@{}lll@{\hskip 2pt}l@{\hskip 3pt}l@{\hskip 3pt}r@{\hskip 2pt}r@{}}
    \toprule
    \textbf{Variant} & \textbf{Frmwk} & \textbf{Store} & \textbf{Components} & \textbf{Integr.} & \textbf{LoC} & \textbf{Files} \\
    \midrule
    A: FAQ & Express & JSON & FAQ loader, chat handler & API call & 141 & 2 \\
    B: JSON & Express & --- & Zod schema, retry, fallback & JSON mode & 472 & 4 \\
    C: Tools & Express & SQLite & Tool registry, agent loop, store & Func.\ call & 827 & 6 \\
    \bottomrule
  \end{tabular}
  \par\vspace{10pt}
  {\raggedright\scriptsize Prompts: A = ``answer product questions from a FAQ''; B = ``return structured JSON (intent, confidence, entities) with schema validation''; C = ``agent with tool access: search\_kb(), get\_account(), escalate\_to\_human()''. Each variant developed independently with no shared context.\par}
\end{table}

The three variants fall along a spectrum of prompt specificity. Variant~A names a task (``answer questions from a FAQ''). Variant~B specifies an output contract (``return JSON with fields: intent, confidence, entities''). Variant~C declares capabilities (``you have access to: search\_kb(), get\_account(), escalate()''). As the prompt grows more specific, the line between prompting and architecture blurs. A capability declaration is, in practice, an interface definition. An output schema is a data contract. This blurring is not surprising, but the central observation. The prompt becomes the architecture specification, and the generated infrastructure follows from it. When agents compose prompts on their own (as in coordinated sub-agent workflows), this coupling operates without human review.

What distinguishes agent-made from human-made architectural decisions? Three properties stand out. \textit{Scale.} Framework, database, authentication, and deployment are selected in a single interaction, packaged as a unit rather than as separately reviewable choices. \textit{Speed.} Decisions that teams deliberate over for days occur within seconds, outpacing any review process. \textit{Opacity.} Choices remain buried in generated code with no ADRs, no design documents, no recorded rationale. The opacity of these decisions has a further consequence. When the systems that agents build are themselves LLM-integrated, the prompt's influence extends beyond behavior and determines the infrastructure the system requires. 

% Architectural components per prompt variant (A, B, C)
% Full-width TikZ figure, black-and-white friendly
% Included via \input{figures/fig-versions.tex}

\begin{figure*}[t!]
\centering
\vspace{-2mm}
\begin{tikzpicture}[
    comp/.style={draw, thick, minimum width=3.8cm, minimum height=0.55cm,
                 align=center, font=\scriptsize},
    lbl/.style={font=\small, align=center},
    metric/.style={font=\tiny, text=black!70},
]

% --- Column positions ---
\def\colA{0}
\def\colB{5.2}
\def\colC{10.4}

% ===================== Variant A =====================
\node[lbl] at (\colA, 3.6) {Variant A: FAQ};
\node[metric] at (\colA, 3.1) {141 LoC, 2 files};

\node[comp, fill=black!5]  (a1) at (\colA, 2.3) {Express Server};
\node[comp, fill=black!5]  (a2) at (\colA, 1.5) {FAQ Loader};
\node[comp, fill=black!10] (a3) at (\colA, 0.7) {Chat Handler};
\node[comp, fill=black!10] (a4) at (\colA, -0.1) {LLM API Call};

\draw[->, thick] (a1) -- (a2);
\draw[->, thick] (a2) -- (a3);
\draw[->, thick] (a3) -- (a4);

% ===================== Variant B =====================
\node[lbl] at (\colB, 3.6) {Variant B: Structured JSON};
\node[metric] at (\colB, 3.1) {472 LoC, 4 files};

\node[comp, fill=black!5]  (b1) at (\colB, 2.3) {Express Server};
\node[comp, fill=black!15] (b2) at (\colB, 1.5) {Zod Schema};
\node[comp, fill=black!15] (b3) at (\colB, 0.7) {JSON Mode Call};
\node[comp, fill=black!20] (b4) at (\colB, -0.1) {Retry Handler};
\node[comp, fill=black!20] (b5) at (\colB, -0.9) {Fallback Generator};

\draw[->, thick] (b1) -- (b2);
\draw[->, thick] (b2) -- (b3);
\draw[->, thick] (b3) -- (b4);
\draw[->, thick] (b4) -- (b5);

% ===================== Variant C =====================
\node[lbl] at (\colC, 3.6) {Variant C: Tool Access};
\node[metric] at (\colC, 3.1) {827 LoC, 6 files};

\node[comp, fill=black!5]  (c1) at (\colC, 2.3) {Express Server};
\node[comp, fill=black!20] (c2) at (\colC, 1.5) {Tool Registry};
\node[comp, fill=black!25] (c3) at (\colC, 0.7) {Agent Loop};
\node[comp, fill=black!25] (c4) at (\colC, -0.1) {Function Calling};
\node[comp, fill=black!30] (c5) at (\colC, -0.9) {SQLite State Store};

\draw[->, thick] (c1) -- (c2);
\draw[->, thick] (c2) -- (c3);
\draw[->, thick] (c3) -- (c4);
\draw[->, thick] (c4) -- (c5);

% --- Arrows between columns ---
\draw[->, thick, dashed, black!50] (a3.east) -- (b3.west)
    node[midway, above, font=\tiny] {+schema};
\draw[->, thick, dashed, black!50] (b3.east) -- (c3.west)
    node[midway, above, font=\tiny] {+tools};

\end{tikzpicture}
\vspace{-2mm}
\caption{Architectural components introduced by each prompt variant. Darker shading indicates components added by increasing prompt specificity. Variant~A passes FAQ context to the LLM and returns free-form text; Variant~B adds Zod validation, JSON Schema mode, and retry logic; Variant~C adds a tool registry, agent loop, and SQLite state management.}
\label{fig:versions}
\end{figure*}
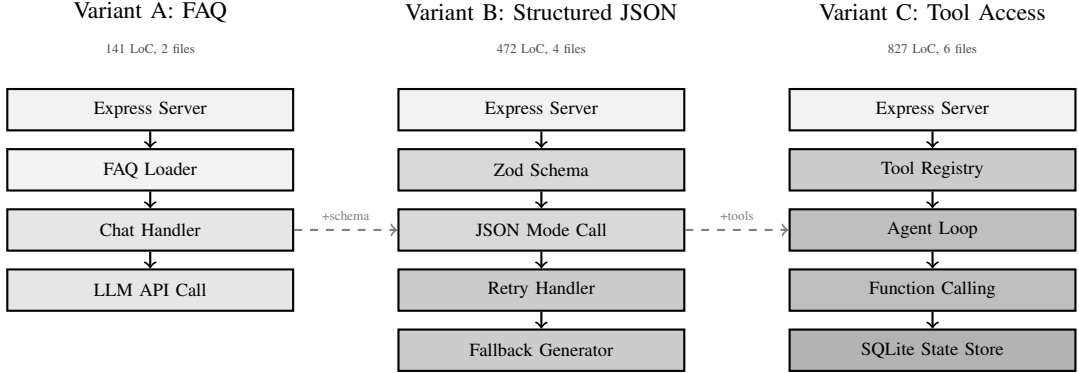

%% ============================================================
\section{Prompt-Architecture Coupling}
\label{sec:coupling}

Agents make architectural decisions at scale, at speed and without transparency. The problem compounds when the systems they build are themselves LLM-integrated, a common situation in 2026. In such systems, prompt design determines which infrastructure components are needed. 

Recent systems research shows that prompt design characteristics, particularly input length and structure, directly determine infrastructure requirements. Longer prompts increase token-processing overhead, forcing architectural choices regarding caching strategy and GPU batching~\cite{b29}. This coupling between prompt design and infrastructure is particularly acute in multi-agent systems where prompts are generated dynamically.

\subsection{The Problem of Implicit Coupling}

The illustrative case study makes the point concretely. Variant~A uses a free-form prompt and results in a flat architecture with 2~files and 141~LoC. Adding a structured output requirement (Variant~B) forces the system to acquire a Zod schema, a retry handler, and a fallback generator, bringing the total to 4~files and 472~LoC. Declaring tool access (Variant~C) pushes further by introducing a tool registry, an agent loop, and an SQLite state store, yielding 6~files and 827~LoC. The same task, expressed through three different prompts, produces systems with different component structures, dependency graphs and failure modes (Fig.~\ref{fig:versions}). What appears to be a minor prompt adjustment can create entirely new, unplanned infrastructure. When an AI agent makes prompt choices on the developer's behalf, the architectural consequences can accumulate unnoticed.

\subsection{Six Recurring Couplings Patterns}

Abstractions in LangChain~\cite{b8}, LlamaIndex~\cite{b9}, and provider documentation~\cite{b30,b31} were traced, following each prompt-level feature to the infrastructure it requires. Six patterns emerged across three categories (Table~\ref{tab:patterns}). Each is classified as either \textit{contingent} (likely to weaken as models gain native capabilities, e.g. native JSON output) or \textit{fundamental} (logically necessary regardless of model capability, e.g., tool execution always requires orchestration). The catalog is not exhaustive but covers the most frequently observed couplings.

\begin{table*}[t!]
  \caption{\small Prompt-Architecture Coupling Patterns}
  \label{tab:patterns}
  \centering
  \footnotesize
  \setlength{\tabcolsep}{4pt}
  \renewcommand{\arraystretch}{1.3}
  \begin{tabular}{@{}lllll@{}}
    \toprule
    \textbf{Pattern} & \textbf{Trigger} & \textbf{Infrastructure Required} & \textbf{Quality Impact} & \textbf{Type} \\
    \midrule
    \multicolumn{5}{@{}l}{\textit{Constraint patterns}} \\[2pt]
    1A: Structured Output~\cite{b8,b30} & JSON schema & Parser, validator, retry, fallback & Interoperability, robustness & Contingent \\
    1B: Few-Shot~\cite{b9} & Dynamic examples & Embedding, vector store, curator & Capability, grounding & Contingent \\
    \midrule
    \multicolumn{5}{@{}l}{\textit{Capability patterns}} \\[2pt]
    2A: Function Calling~\cite{b30,b31} & Tool signatures & Router, validator, error handler, agent loop & Extensibility, attack surface & Fundamental \\
    2B: ReAct Reasoning~\cite{b8,b33} & CoT with tools & State machine, validator, timeout handler & Autonomy, testability & Fundamental \\
    \midrule
    \multicolumn{5}{@{}l}{\textit{Context patterns}} \\[2pt]
    3A: RAG~\cite{b9,b8} & Bounded context & Ingest, chunk, embed, vector store, ranker & Accuracy, infrastructure cost & Contingent \\
    3B: Context Reduction~\cite{b9} & Token budget & Summarizer, filter, extractor & Cost, information loss & Contingent \\
    \bottomrule
  \end{tabular}
  \par\vspace{10pt}
  {\footnotesize Contingent couplings may weaken as models gain native capabilities; fundamental couplings are logically necessary regardless of model capability. Composition multiplies effects super-linearly.\par}
\end{table*}

Constraining the form of output introduces the first class of coupling. A prompt requiring structured responses (``return JSON with fields: intent, confidence, entities'') forces a parser, schema validator, retry handler, and fallback generator into the system. LangChain's \texttt{with\_structured\_output()}~\cite{b8} and OpenAI's strict mode~\cite{b30} wrap exactly this pipeline. In the case study, Variant~B acquired a Zod schema, exponential-backoff retry, and fallback, three components absent from Variant~A, solely because the prompt demanded structured output. Dynamic few-shot selection creates a parallel dependency. When a prompt selects examples at query time rather than hard-coding them, the system needs an embedding model, a vector store for the example bank, and an example curator~\cite{b9}. Both couplings are contingent. As models gain native JSON guarantees or larger context windows, the validation and retrieval stacks may shrink.

Declaring what a model can do introduces a different, more durable kind of coupling. Typed tool signatures (\texttt{search\_kb(query:~string)}) create a service-oriented architecture with a function router, argument validator, error handler, and response loop~\cite{b30,b31}. Variant~C's three tool declarations produced a tool registry, an agent loop (ten iterations), and an SQLite state store, none of which exist in~A or~B. Each additional tool widens the attack surface~\cite{b32}. ReAct-style reasoning~\cite{b33} layers a state machine on top, tracking reasoning steps with per-step validation, partial failure handling, and timeout enforcement (LangGraph's \texttt{StateGraph}~\cite{b8} is a direct implementation). Both couplings are fundamental. Tool execution will always require orchestration, and interleaved reasoning steps resist unit testing~\cite{b12} regardless of how capable models become.

The remaining two patterns control which information reaches the model. Constraining answers to retrieved context requires an ingestion, chunking, embedding, and ranking pipeline~\cite{b34,b35}. Variant~A's FAQ injection is a minimal static analogue. At scale, the relevance ranking remains valuable even as context windows expand. Token budget constraints add a complementary pre-processing stage, summarization, filtering, extraction~\cite{b9}, before the prompt reaches the model. Privacy filtering, cost control, and batch processing keep this pattern relevant despite growing context limits. LlamaIndex documents several node-level transformations for both retrieval and reduction. The trade-off in each case is infrastructure complexity versus information quality.

Across all six patterns, the direction is the same: the prompt drives the architecture. The couplings raise practical questions, in particular, how teams should review prompt-level decisions and what tooling would make the resulting architectural choice visible. The following section addresses both.

% Composition of prompt-architecture coupling patterns
% Full-width TikZ figure, black-and-white friendly
% Included via \input{figures/fig-composition.tex}

\begin{figure*}[t]
\centering
\begin{tikzpicture}[
    block/.style={draw, thick, minimum height=0.6cm, align=center, font=\scriptsize},
    pattern/.style={draw, thick, minimum width=2.2cm,
                    minimum height=0.6cm, align=center, font=\scriptsize},
    crosscut/.style={draw, thick, dashed, minimum width=13.5cm,
                     minimum height=0.45cm, align=center, font=\scriptsize,
                     fill=black!8},
    arr/.style={->, thick, >=latex, shorten >=3pt, shorten <=3pt},
    lbl/.style={font=\tiny, text=black!60},
]

% --- Main pipeline (left to right) ---
\node[block, minimum width=1.8cm, fill=black!5] (user) at (0, 0) {User Query};

% Pattern 3A: RAG
\node[pattern, fill=black!12] (chunk) at (3.2, 0) {Chunking};
\node[pattern, fill=black!12] (embed) at (3.2, -1.0) {Embedding};
\node[pattern, fill=black!12] (rank)  at (3.2, -2.0) {Ranking};
\node[lbl] at (3.2, 0.6) {3A: RAG Pipeline};

% Pattern 3B: Context reduction
\node[pattern, fill=black!18] (summ) at (6.4, 0) {Summarization};
\node[lbl] at (6.4, 0.6) {3B: Context Reduction};

% Pattern 1A: Structured output
\node[pattern, fill=black!15] (llm) at (9.2, 0) {LLM + Schema};
\node[pattern, fill=black!15] (valid) at (9.2, -1.0) {Validator};
\node[pattern, fill=black!15] (retry) at (9.2, -2.0) {Retry / Fallback};
\node[lbl] at (9.2, 0.6) {1A: Structured Output};

% Pattern 2A: Function calling
\node[pattern, fill=black!22] (router) at (12.4, 0) {Tool Router};
\node[pattern, fill=black!22] (exec)   at (12.4, -1.0) {Tool Executor};
\node[pattern, fill=black!22] (state)  at (12.4, -2.0) {State Store};
\node[lbl] at (12.4, 0.6) {2A: Function Calling};

% --- Main pipeline flow (left to right at top level) ---
\draw[arr] (user.east) -- (chunk.west);
\draw[arr] (chunk.east) -- (summ.west);
\draw[arr] (summ.east) -- (llm.west);
\draw[arr] (llm.east) -- (router.west);

% --- Internal vertical flows within each pattern ---
\draw[arr] (chunk) -- (embed);
\draw[arr] (embed) -- (rank);
\draw[arr] (llm) -- (valid);
\draw[arr] (valid) -- (retry);
\draw[arr] (router) -- (exec);
\draw[arr] (exec) -- (state);

% --- Cross-cutting concerns (dashed bars below) ---
\node[crosscut] at (7.8, -3.2) {Authentication};
\node[crosscut] at (7.8, -3.9) {Rate Limiting};
\node[crosscut] at (7.8, -4.6) {Logging \& Monitoring};

\end{tikzpicture}
\caption{Composition of coupling patterns in a single system (RAG chatbot with tool access). Pattern IDs match the catalog in Section~\ref{sec:coupling}. Cross-cutting concerns (dashed) span multiple patterns and add infrastructure no single pattern accounts for on its own.}
\label{fig:composition}
\end{figure*}
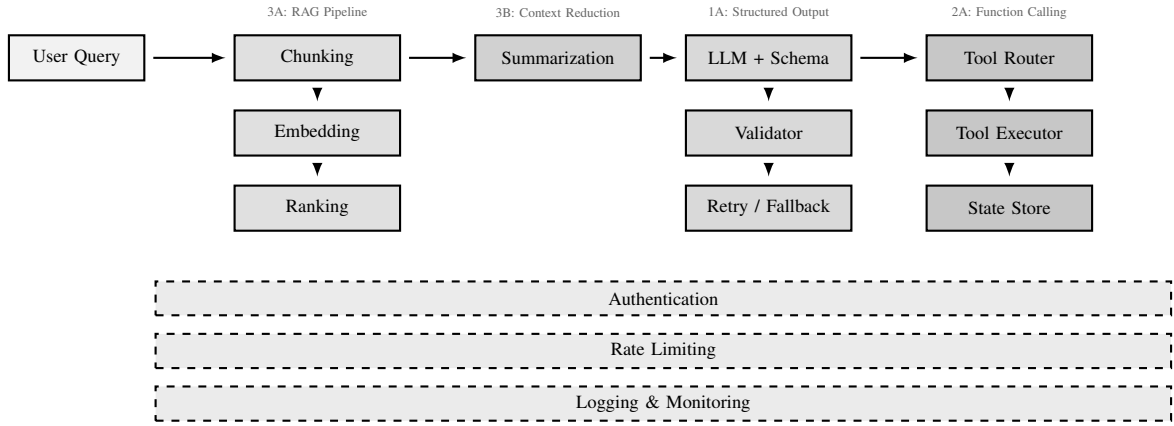

%% ============================================================
\section{Discussion}
\label{sec:discussion}

\subsection{How Agentic Coding Is Changing Architecture}

Agent-scaffolded projects are converging on a narrow set of stacks. Bolt.new, Lovable, and v0~\cite{b36} all default to React, TypeScript, and Tailwind. This simplifies knowledge transfer but concentrates vulnerability exposure~\cite{b25}. At the same time, every decision is implicit. A developer typing ``a todo app with auth'' receives database, authentication, and deployment choices already made, with no rationale on record. Kravchuk-Kirilyuk~\textit{et~al.}~\cite{b17} show that such code appears modular on the surface while hiding deeper coupling, and developers knowingly skip validation~\cite{b16}.

The resulting speed-review gap is acute. Agents scaffold systems in minutes; teams need hours or days to audit them. Tool-use patterns (2A, 2B) widen the attack surface through indirect prompt injection~\cite{b32}. MCP~\cite{b26} standardizes tool integration but also standardizes attack vectors. The \texttt{AGENTS.md} convention~\cite{b27} is a first guardrail, but securing composed patterns remains open when multiple tool-use patterns combine (Fig.~\ref{fig:composition}).

\subsection{Implications for Practice}

Prompt specifications are architectural artifacts and belong in architectural review. An impact statement (e.g., ``Structured JSON output adds validator, retry handler, fallback; +330~LoC, two new failure modes'') would make the coupling explicit before code generation begins. Prompt choices and agent decisions also belong in ADRs. Existing tools (Claude Code hooks, \texttt{.cursorrules}, \texttt{AGENTS.md}) constrain reactively. Proactive guidance that flags consequences before generation is still missing. The 5.9$\times$ code growth (141$\rightarrow$827~LoC) and 3$\times$ file increase in the case study (Table~\ref{tab:mechanisms}) suggest that complexity thresholds, baselines that trigger review when exceeded, would help teams catch uncontrolled scaffolding early.

\subsection{Toward Architecture-Aware AI-Assisted Development}

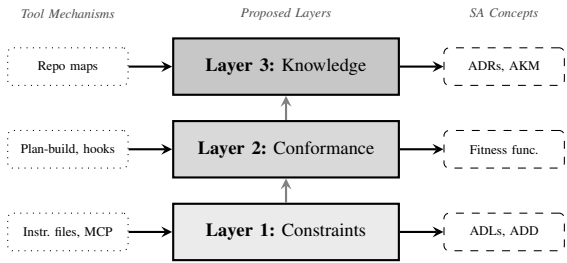
\begin{figure}[t]
\centering
\begin{tikzpicture}[
    layer/.style={draw, thick, minimum width=3.0cm, minimum height=0.75cm, align=center, font=\scriptsize},
    sabox/.style={draw, thin, dashed, rounded corners=2pt, minimum width=1.6cm, minimum height=0.5cm, align=center, font=\tiny},
    toolbox/.style={draw, thin, dotted, rounded corners=2pt, minimum width=1.6cm, minimum height=0.5cm, align=center, font=\tiny},
    arrow/.style={->, thick, >=stealth},
    lbl/.style={font=\tiny\itshape, text=black!70},
]

% Layers (bottom to top)
\node[layer, fill=black!8] (L1) at (0,0) {\textbf{Layer 1:} Constraints};
\node[layer, fill=black!15] (L2) at (0,1.1) {\textbf{Layer 2:} Conformance};
\node[layer, fill=black!22] (L3) at (0,2.2) {\textbf{Layer 3:} Knowledge};

% SA concepts (right side)
\node[sabox] (sa1) at (2.9,0) {ADLs, ADD};
\node[sabox] (sa2) at (2.9,1.1) {Fitness func.};
\node[sabox] (sa3) at (2.9,2.2) {ADRs, AKM};

% Tool mechanisms (left side)
\node[toolbox] (t1) at (-2.9,0) {Instr.\ files, MCP};
\node[toolbox] (t2) at (-2.9,1.1) {Plan-build, hooks};
\node[toolbox] (t3) at (-2.9,2.2) {Repo maps};

% Arrows: layers to SA concepts
\draw[arrow] (L1.east) -- (sa1.west);
\draw[arrow] (L2.east) -- (sa2.west);
\draw[arrow] (L3.east) -- (sa3.west);

% Arrows: tool mechanisms to layers
\draw[arrow] (t1.east) -- (L1.west);
\draw[arrow] (t2.east) -- (L2.west);
\draw[arrow] (t3.east) -- (L3.west);

% Inter-layer arrows
\draw[arrow, black!50] (L1.north) -- (L2.south);
\draw[arrow, black!50] (L2.north) -- (L3.south);

% Column labels
\node[lbl] at (-2.9, 2.9) {Tool Mechanisms};
\node[lbl] at (0, 2.9) {Proposed Layers};
\node[lbl] at (2.9, 2.9) {SA Concepts};

\end{tikzpicture}
\caption{Three-layer framework for architecture-aware AI-assisted development. Each layer maps tool mechanisms (left) to SA concepts (right).}
\label{fig:vision}
\end{figure}

A common gap connects both phenomena. Agents make architectural decisions, but no feedback loop ties those decisions to established architectural knowledge. This paper sketches a three-layer framework (Fig.~\ref{fig:vision}) mapping existing tool mechanisms to software architecture concepts.

% Three-layer vision for architecture-aware vibe coding
% Standalone TikZ figure — included via \input{} in main.tex
% Designed for single-column width, black-and-white

The first layer, \emph{constraints}, specifies what the agent may and may not do. Instruction files (\texttt{AGENTS.md}, \texttt{.cursorrules}) and MCP server configurations~\cite{b26} already play this role informally. Architecture Description Languages (ADLs) and Attribute-Driven Design (ADD)~\cite{b25} formalize the same concern. The second layer, \emph{conformance}, checks generated code against those constraints. Plan-build workflows (where the agent proposes before executing) and post-generation hooks are analogous to fitness functions in evolutionary architecture. The third layer, \emph{knowledge}, feeds architectural context back to the agent. Repository maps and context files serve this purpose today; ADRs and Architectural Knowledge Management (AKM) do so in the SA literature. Concretely, a conformance layer could compare the dependency graph of generated code against declared constraints (e.g., ''no new database dependencies without review'') and flag violations before they propagate. The knowledge layer could extract architectural decisions from agent reasoning traces and persist them as ADRs, closing the documentation gap between generation speed and review capacity.

Falc{\~a}o~\textit{et~al.}~\cite{b20} envision a trajectory toward self-coding systems with runtime autonomy. The framework here is complementary, targeting development-time governance with a human in the loop rather than runtime self-adaptation. Whether the two converge as agents grow more autonomous is an open question.

\subsection{Threats to Validity}

Three validity concerns bound the claims made in this paper and shape the next steps. For \textit{construct validity}, ``architectural decision'' is defined following Bass~\textit{et~al.}~\cite{b25}, and ``vibe architecting'' is framed as a deliberate extension of Karpathy's ``vibe coding''~\cite{b1}. The six coupling patterns are analytical constructs proposed for community discussion, not yet empirically validated.

For \textit{internal validity}, the prompts encode architectural expectations at varying specificity; future work should test with minimally prescriptive prompts to isolate what the agent decides unprompted. Quality impact assessments in Table~\ref{tab:patterns} are analytical, not empirical.

For \textit{external validity}, the tool survey reflects agentic coding as of early 2026, a space where new tools appear regularly. The six coupling patterns draw on three frameworks and additional patterns almost certainly exist beyond this set. Each prompt in the illustrative example was run once, with one tool and one LLM, so replicating across agents, models, and developer populations would test whether the observed architectural divergence generalizes.

\subsection{Research Agenda}
\label{sec:agenda}

Five directions follow from this analysis. First, cross-agent replication. The case study varied prompts while holding the agent constant. Do Claude Code~\cite{b2}, Cursor~\cite{b3}, and Devin~\cite{b4} produce substantially different architectures for the same specification? If so, agent selection is itself an architectural decision requiring governance.

Second, architectural footprint metrics. LoC and file count (Table~\ref{tab:mechanisms}) are coarse. Richer metrics capturing component count, dependency depth, and new failure modes would let teams flag prompt changes exceeding complexity thresholds.

Third, proactive governance tooling. Existing guardrails (hooks, \texttt{.cursorrules}, \texttt{AGENTS.md}~\cite{b27}) constrain reactively. A tool previewing architectural impact (``this adds vector database, embedding pipeline, async job queue'') before code generation would address the review-speed gap.

Fourth, automated ADR generation. The rationale behind architectural choices exists implicitly in the agent's reasoning trace; extracting it into structured ADRs would close the documentation gap.

Fifth, pattern composition analysis. Fig.~\ref{fig:composition} shows that coupling patterns compose. A RAG chatbot with tool access pulls in retrieval, function calling, and structured output simultaneously. Whether cross-cutting concerns (authentication, rate limiting, logging) scale additively or super-linearly with pattern composition remains an open question.

%% ============================================================
\section{Conclusion}
\label{sec:conclusion}

AI coding agents make architectural decisions through five identifiable mechanisms, and prompt-architecture coupling produces six recurring patterns where natural-language instructions determine infrastructure. Varying the prompt wording alone yielded structurally different systems in the illustrative example, from 141 to 827 lines of code and two to six files for the same task.

Vibe coding becomes vibe architecting the moment a prompt determines system structure. The question for the architecture community is whether these prompt-level decisions deserve the same governance as design-level ones. This paper argues that they do. The review practices, decision records, and tooling to make that governance possible do not yet exist. Building them is the next step.

Vibe architecting is a research program, not a single contribution. The five directions in Section~\ref{sec:agenda} each connect to existing work in software architecture while raising questions specific to the AI-assisted setting. As agents grow more capable and autonomous, the governance gap will only widen. These starting points are intended to invite both empirical follow-ups and tool-building efforts from the community.

\bibliographystyle{IEEEtran}
\bibliography{refs}

\end{document}